\documentstyle[twoside,aaspp4,psfig]{article}
\newcommand{\Msolar}{\mbox{\,$M_{\odot}$\/}}          % solar mass
\newcommand{\Mearth}{\mbox{\,$M_{\oplus}$\/}}         % earth mass
\newcommand{\HII}{\mbox{H\,{\footnotesize II}}}       % HII properly spaced
\newcommand{\magnit}[2]{\mbox{$ #1^m \!\!\!.\!\,\, #2$}}   % magnitudes
\newcommand{\magap}[1]{\mbox{$ #1^m$}}                     % mag w/o d.p.
\newcommand{\filter}[1]{\mbox{\it #1\/}}              % filter in italics
\newcommand{\oversim}[2]{\lower0.5ex\vbox{\baselineskip=0pt\lineskip=0.2ex
     \ialign{$\mathsurround=0pt #1\hfil##\hfil$\crcr#2\crcr\sim\crcr}}} 
\newcommand{\simless} {\mbox{$\mathrel{\mathpalette\oversim<}$}} % <~ sign
\newcommand{\etal}{\mbox{\hbox{\it et\,al.}}}         % et al. in italics
\newcommand{\eg}{\mbox{\hbox{\it e.g.},}}             % e.g. in italics
\newcommand{\idest}{\mbox{\hbox{\it i.e.},}}          % i.e. in italics
\newcommand{\cf}{\mbox{\hbox{\it cf.}}}               % cf. in italics
\begin{document}

\title{High-resolution near-infrared imaging of the Orion~114-426 
silhouette disk\footnote{%
To appear in the special NICMOS/STIS ERO edition of the
{\it Astrophysical Journal (Letters),} January 1998}
}

\author{%
Mark J. McCaughrean\footnote{%
Max-Planck-Institut f\"ur Radioastronomie,
Auf dem H\"ugel 69,
53121 Bonn,
Germany;
mjm@mpifr-bonn.mpg.de},
% \mbox{}\hspace{21pt}E-mail: mjm@mpifr-bonn.mpg.de},
%
Hua Chen\footnote{%
Steward Observatory,
University of Arizona,
933 N\@. Cherry Ave.,
Tucson, AZ~85721} 
John Bally\footnote{%
Center for Astrophysics and Space Astronomy,
University of Colorado,
Boulder, CO~80309--089}
% \mbox{}\hspace{21pt}Boulder, CO~80309--089}
%
Ed Erickson\footnote{%
NASA Ames Research Center,
MS 245--6,
Moffett Field, CA~94035--1000}
Rodger Thompson\footnotemark[3],
Marcia Rieke\footnotemark[3],
Glenn Schneider\footnotemark[3],
Susan Stolovy\footnotemark[3], and
Erick Young\footnotemark[3]
}
\begin{abstract}
\rightskip 0pt
We present the first high-resolution near-infrared images of the edge-on 
silhouette circumstellar disk, Orion~114-426, made using NICMOS on the 
{\it Hubble Space Telescope}. Images taken against the bright nebular 
background of the ionized hydrogen Pa$\alpha$ line at 1.87\micron{} 
show the major axis of the disk to be approximately 20\% smaller than 
at 0.6\micron, from which we deduce the structure of the edge of the
disk. Continuum images of diffuse polar lobes above and below the plane 
of the disk show a morphology and evolution with wavelength consistent 
with predictions for reflection nebulae in a diffuse envelope with large 
polar cavities, surrounding a thin, massless, Keplerian disk, centered
on an otherwise hidden central star. We make use of our observations and 
reasonable assumptions about the underlying disk structure to show that 
the disk mass is at least 10\Mearth{} and plausibly 
$\geq 5\times 10^{-4}$\Msolar.
\end{abstract}

Subject headings: 
accretion, accretion disks --- 
circumstellar matter ---
ISM: individual (Orion Nebula) --- 
stars: formation, pre-main sequence ---
infrared: ISM: continuum, lines and bands

\section{Introduction}
The discovery of a family of circumstellar disks seen as dark silhouettes in
projection against the Orion Nebula using the {\it Hubble Space Telescope\/}
provided strong confirmation of the disk paradigm of star formation (O'Dell 
\etal{} 1993; O'Dell \& Wen 1994; McCaughrean \& O'Dell 1996 [MO96]). The 
disks range in diameter from 50--1000\,AU, and thus the $\sim$50\,AU (0.1 
arcsec at 450\,pc) resolution of the HST observations was sufficient to 
examine their structure directly at optical wavelengths. An important 
finding was that the radial surface density profiles appear to be abruptly 
truncated at some outer radius, perhaps due to external effects from the 
surrounding \HII{} region and dense cluster (MO96), and more detailed 
examination of this transition zone should lead to a greater understanding 
of the evolution of disks in harsh environments.

The discovery images were obtained over a relatively narrow wavelength range 
(5007--6585\AA), and further insight should be possible through HST 
observations at shorter and longer wavelengths. In the blue/near-UV 
($\sim$\,2000--4000\AA), the spatial resolution approaches $\sim$15\,AU, 
while increased dust opacity at these wavelengths should also allow more 
tenuous structures to be traced to larger radii. Conversely, the considerable 
{\em reduction\/} in dust opacity at near-IR wavelengths should allow us to 
trace structures to smaller radii, albeit with commensurately poorer spatial 
resolution. Consequently, we are conducting follow-up HST studies from the 
near-UV to near-IR (0.3--2.5\micron), and in the present paper, we report 
preliminary near-IR observations using NICMOS of one silhouette disk, 
Orion~114-426. The largest of the sample at $\sim$1000\,AU diameter, 
this disk is seen near edge-on, and while the central star is not directly 
visible at optical wavelengths, its presence is betrayed by two polar 
nebulosities believed to be illuminated by it. 

\section{Observations}
A comprehensive General Observer program (McCaughrean \etal: GO\,7367) 
studying the Orion silhouette disks with NICMOS, STIS, and WFPC2 is
being carried out during HST Cycle~7. Early Release Observations using 
NICMOS were subsequently proposed by the Instrument Development Team 
(Erickson \etal: SM2/ERO\,7114) for scientific verification and media use. 
Due to this overlap, the ERO data were reduced and analysed collaboratively, 
resulting in studies of 114-426 (presented here) and of the 182-413/183-419 
field (Chen \etal{} 1998).

NICMOS observations of the 114-426 field were obtained on 19 April 1997 
during the Servicing Mission Orbital Verification following installation 
in the HST\@. Images were taken through broad-band, narrow-band, and 
polarimetric filters between 1 and 2.1\micron{} as summarized in Table~1. 
Data reduction combined standard ground-based near-IR imaging techniques 
with parts of the NICMOS calibration pipeline. Multiple read-outs combined 
with multiple positions on the sky were used to reject cosmic-ray events; 
electronic offsets were removed with on-orbit dark images; quantum 
efficiency variations were removed with flat fields taken on-orbit
where possible, otherwise from ground tests. Finally, mosaics were 
made registering the multiple images using stars or HST pointing information.
Detailed photometric calibration was not attempted, but ground-based 
near-IR magnitudes for stars in the field were used to calibrate within 
$\pm$\magnit{0}{2}. 

Despite integration times significantly shorter than those planned for 
the GO program, important preliminary results were nevertheless obtained
from the narrow-band imaging against the bright Pa$\alpha$ background at 
1.87\micron, broad-band imaging at 1.1 and 1.6\micron, and the polarization 
imaging at 2.0\micron. The three polarizer position images were combined to 
form a 2\micron{} continuum image, but due to remaining uncertainties in 
the correct analysis techniques for NICMOS polarimetry and incomplete 
on-orbit polarization calibration, the polarization results themselves 
are deferred to a future paper. The remaining narrow-band images did not 
provide useful additional information and are not further discussed. 

\section{Results}
\subsection{Narrow-band imaging in Pa$\alpha$}
The highest S/N images of the silhouettes obtained by MO96 were through a 
narrow-band H$\alpha$ ($\lambda$6565\AA) filter, admitting the full 
emission line flux from the bright Orion Nebula \HII{} region, while 
minimizing continuum emission from the central stars, or in the case 
of 114-426, its polar lobes. The brightest near-IR counterpart is the 
Pa$\alpha$ line at 1.87\micron, which cannot be detected from the ground 
due to atmospheric absorption. For typical \HII{} region ionization 
parameters (10$^4$\,K, 10$^4$\,cm$^{-3}$, Case B) and $A_V$$\sim$\magap{1} 
foreground to the nebula, the detected photon flux at Pa$\alpha$ should 
be $\sim$\,60\% of that at H$\alpha$: the brightest equivalent line available 
to ground-based observers (Br$\gamma$ at 2.16\micron) would be a further 
factor of ten fainter (Osterbrock 1989).

The Pa$\alpha$ 1.87\micron{} image of 114-426 is shown in Figure~1 with
the H$\alpha$ ($\lambda$6565\AA) image from MO96\@. The S/N in the P$\alpha$ 
image is poor ($\simless$5:1) since the integration time was short (288 sec), 
and the NIC1 image scale of 0.0432 arcsec/pixel over-resolved the 0.19 arcsec 
FWHM diffraction-limited resolution of the telescope at 1.87\micron. 
% Also, the disk lies on a bright and noisy
% region of the detector that is an artifact of poor flat-fielding: no 
% on-orbit flat was available for the F187N filter at time of writing. 
Nevertheless, the silhouette is clearly detected, allowing a preliminary 
measurement of its size. The data were binned by a factor of two to
better match the appropriate pixel size (\idest{} 2 pixels per FWHM) and
then averaged across the minor axis. The resulting 1D major axis profile 
had high enough S/N to show the two ends of the disk as sharp dips 
separated by 1.8 arcsec. As discussed in detail by MO96, the apparent size 
and structure of a silhouette disk is a convolution of its real form with 
the instrumental point spread function, and following MO96, we adjusted the 
parameters of a model edge-on disk convolved with a model HST+NICMOS PSF 
calculated using the Tiny\,Tim software (Krist \& Hook 1997) until the major 
axis length was reproduced. The resulting best-fit model disk has a major 
axis size of $\simeq$\,800\,AU at 1.87\micron, $\sim$20\% less than the 
1012\,AU measured at 0.6\micron{} (MO96). The same procedure used on the
2\micron{} continuum image (Section~\ref{sec:continuum}) yielded the same
result to within 5\%. Finally, we verified the overall procedure by 
degrading the high S/N [O\,III] image from MO96 to the same spatial
resolution and S/N as the Pa$\alpha$ image, then performing the same
fitting process, before retrieving the correct size for the disk at optical
wavelengths. 

% The implications of this finding for the disk
% structure are discussed in Section~\ref{sec:structure}.

\subsection{Continuum imaging of the polar lobes} \label{sec:continuum}
The optical continuum image of 114-426 showed faint polar lobes, interpreted 
as reflection nebulae of tenuous dust above and below the plane of the disk, 
illuminated by the otherwise unseen central star (MO96). Similar reflection
nebulae are seen above and below the plane of an edge-on disk in the HH\,30
system (Burrows \etal{} 1996). The wavelength dependent morphology and 
polarization structure of the lobes in 114-426 should allow us to probe 
the underlying form of the disk, the geometry of polar cavities, the grain 
size, and scattering function. 

The near-IR broad-band continuum images, along with the F547M image from MO96, 
are shown in Figure~2 in grayscale and contour forms. As the wavelength 
increases, three effects are seen. First, the initially fainter SE polar lobe 
increases in brightness until it equals then outshines the intensity of the 
NW lobe. The peak intensities in the lobes (after background subtraction) are 
in the ratios 7.3:1, 2.2:1, 1.2:1, and 0.85:1 for the NW:SE lobes, at 0.57, 
1.1, 1.6, and 2.0\micron{} respectively. Second, the nebulae move closer
together, as the reduced extinction allows us to probe closer to the disk 
midplane: the separations of the peak pixels in the two lobes
are at 0.64, 0.43, 0.40, and 0.32 arcsec at 0.57, 1.1, 1.6, and 2.0\micron{} 
respectively. Third, the nebulae appear to flatten from conical to slab-like. 

These features can be compared to model disks and envelopes (\eg{} Lazareff, 
Pudritz, \& Monin 1990; Whitney \& Hartmann 1992 [WH92], 1993 [WH93]; 
Fischer, Henning, \& Yorke 1996 [FHY96]). The general broad fan shape 
and increasing flatness of the nebulae are best reproduced by model 
SH of FHY96, \idest{} a thin, massless disk (in comparison to the central 
star) in Keplerian rotation, with an envelope and broad polar cavities. 
Models with thick, massive disks have more polar material and result in 
images with too much elongation perpendicular to the disk. The same is
true of models with just a narrow, cylindrical polar hole rather than a
broader, so-called ``streamline'' cavity (WH93).

Since the central star in 114-426 is not seen, the disk must lie 
within a few degrees of edge-on, as the thin disk of the SH model does
not occult the central star unless this condition is met (FHY96). This 
degree of alignment is also argued for on the grounds that the two lobes 
have nearly equal brightness in the near-IR\@. The asymmetry between the 
two lobes in the optical continuum (Figure~2a) can probably explained by 
asymmetries in the outer, more diffuse parts of the envelope, perhaps due 
to external effects in the \HII{} region. 

\section{Discussion}
\subsection{Disk structure} \label{sec:structure}
The disk appears $\sim$\,20\% smaller at 1.87\micron{} than at 0.6\micron,
and thus we are clearly resolving structure in its outer parts. In order 
to understand the implications of these observations, we need to examine
the theoretical expectations. For a thin, massless, Keplerian disk that 
is hydrostatically supported and vertically isothermal (Shakura \& Sunyaev 
1973; Pringle 1981; Lazareff \etal{} 1990), the density $\rho$ as a function 
of radius, $r$, and height above the midplane, $z$, is: 
\begin{equation}
\rho(r,z) = \rho_d \left( \frac{r}{r_d} \right)^{-15/8}
            \exp \left[ -\frac{\pi}{4} 
                 \left( \frac{z}{h(r)} \right)^2 \right]
\label{eq:one}
\end{equation}
where $\rho_d$ is the midplane density at the outer disk radius, $r_d$,
and $h(r)$ is the disk scale-height:
\begin{equation}
h(r) = z_d \left( \frac{r}{r_d} \right)^{9/8} 
\label{eq:two}
\end{equation}
where $z_d$ is the scale-height at $r_d$. The disk surface density is then:
\begin{equation}
\Sigma (r) = 2 \rho_d z_d \left( \frac{r}{r_d} \right)^{-3/4}
\label{eq:three}
\end{equation}

For typical Orion Nebula dust grains (R=5; Cardelli, Clayton, \& Mathis
1989), the extinction at 1.87\micron{} is one sixth of that at 0.6\micron,
and thus achieving the same effective optical depth requires six times
higher column density at the longer wavelength. For a face-on disk with 
unity optical depth at 0.6\micron{} at 506\,AU radius, Eq.\,\ref{eq:three} 
shows that the equivalent optical depth at 1.87\micron{} would occur at 
46\,AU, \idest{} the disk would appear much smaller in the near-IR\@. For 
an edge-on disk, the calculation is harder: we are probing the midplane 
density of the disk, not the surface density, and the line-of-sight through 
the disk at a given ``impact parameter'' (\idest{} the distance off-center) 
integrates over different densities at different radii. Assuming the disk is 
truncated at some outer radius, we can integrate the total column density 
through the midplane as a function of the impact parameter, $a$. We have 
calculated profiles for disks in which the midplane density scales as 
$r^\alpha$, with $\alpha = -1, -2,$ and $-3$ (Figure~3). These values were 
chosen since simple analytical integrals exist in each case, but they also 
closely correspond to plausible disk models: $\alpha = -9/8 \simeq -1$ would 
yield a surface density independent of radius; $\alpha = -15/8 \simeq -2$ 
yields the canonical $\Sigma(r) \propto r^{-3/4}$; $\alpha = -21/8 \sim -3$ 
yields $\Sigma(r) \propto r^{-3/2}$, a commonly assumed density law 
for more massive circumstellar disks (\cf{} Adams, Shu, \& Lada 1988). 

Rough power laws can be fit to the inner section of each curve in Fig.\,3, 
with the column density increasing as $a^\gamma$, and $\gamma \simeq -1/4, 
-1$, and $-2$ corresponding to $\alpha = -1, -2,$ and $-3$ respectively. 
In the canonical case of $\alpha = -2 \simeq -15/8$, a disk with unity
optical depth at 506\,AU at 0.6\micron{} would have the same optical depth 
at 84\,AU at 1.87\micron. Thus, even though an edge-on disk will shrink 
less with wavelength than a face-on one, it is clear that for 114-426, the 
observed decrease in size of only 20\% (\idest{} $-0.1$\,dex) for an increase 
in density of six (\idest{} 0.78\,dex) would require $\gamma\sim -8$, and 
thus $\alpha\sim -9$ and $\Sigma(r)\propto r^{-8}$, quite inconsistent with 
conventional understanding of disk structure. 

The exception comes near the edge of a disk, where the integrated column 
density rises sharply as the path length increases most rapidly. Indeed, 
Fig.\,3 shows that in the $\alpha = -2$ case, the density {\em does\/} 
increase by the required factor of six within 80\% of the outer radius, as
observed. Thus, the observed shift between the optical edge at 506\,AU and 
the infrared edge at 400\,AU could still be consistent with the midplane
density following the canonical power law for a thin, massless disk, {\em 
as long as\/} the disk is physically truncated somewhere near the optical 
edge. As discussed by MO96, it seems plausible that the disk may be
truncated due to physical processes present in the Orion Nebula \HII{}
region and/or Trapezium Cluster, and it may be possible to ascertain
which particular process is responsible from our future HST observations,
which will trace the radial structure of 114-426 over a full order of 
magnitude in limiting column density.

\subsection{The disk mass revisited}
The mass of the 114-426 disk remains unknown: in particular, does it exceed 
the $\sim$0.01\Msolar{} of the minimum mass solar nebula (MMSN), \idest{} is
it massive enough to form a planetary system similar to our own? O'Dell \& 
Wen (1994) and MO96 discuss how the mass of a silhouette disk can be roughly 
estimated, noting that the minimum intensity seen towards it is typically 
$\sim$10\% of that of the background \HII{} region. Assuming this to be 
background emission attenuated by the disk, a line-of-sight column density 
is calculated for each point in the silhouette, and summing over the whole 
disk area, a total mass is estimated. However, virtually all of this ``disk 
light'' is an artifact, as the the background emission is blended into the 
disk by the instrumental PSF (MO96). Thus the usefulness of the ``transmission 
technique'' is diminished, but it at least provides an absolute 
{\em lower-limit\/} mass, an important counterpoint to the results from 
millimeter interferometry which are, as yet, generally provide only 
{\em upper-limit\/} masses for disks in the Trapezium Cluster (Mundy, 
Looney, \& Lada 1995; Lada \etal{} 1996).

MO96 used the transmission technique to estimate lower-limit masses 
for the six silhouette disks based on their optical images. The masses 
were small, typically several orders of magnitude less than the MMSN\@. 
The most massive was 114-426 at 0.002\Msolar, but we must report here 
that an error was made in calculating the area and thus 
mass of 114-426 in MO96: its minimum mass should be revised downwards to 
$2\times 10^{-4}$\Msolar. (The masses for the other disks given in Table~2 
of MO96 remain unaffected by this error).

In principle, the transmission technique could be applied to our new near-IR 
images of 114-426, resulting in a significantly {\em higher\/} estimate of 
the minimum mass. Since the disk is only marginally smaller at 1.87\micron, 
the reduced near-IR extinction should lead to a higher 
mass, on the order of the factor of six in $A_V/A_{1.87\mu{\rm m}}$. However, 
the present images are not good enough to make such estimates: the Pa$\alpha$ 
image has very poor S/N, making a reliable measure of the disk/background 
flux ratio impossible, and while the continuum images have much higher S/N, 
the silhouette is contaminated by the reflection nebulae, making it difficult 
to measure the underlying disk/background flux ratio and the true silhouette 
area. Such calculations must wait until higher-quality Pa$\alpha$ data 
are available. 

In the interim, alternative approaches to estimating the disk mass can be 
taken. The non-detection of the central star in the POL 2\micron{} continuum 
image can be used to set a lower limit on the total column density through 
the disk midplane, depending on the intrinsic brightness of the star. The 
integrated flux from the two reflection nebulae is \filter{K}=\magnit{14}{5}, 
which clearly only represents a fraction of the flux we would see from the 
central star if it had no disk. Some estimate of this fraction can be made
from WH92 and WH93, who tabulate $F/F_{\star}$ (the total flux observed as 
a fraction of the flux that would be seen if the star had no disk or envelope) 
for a variety of models. While $F/F_{\star}$ ranges from 0.01--10\% depending 
on the disk and envelope structure, inclination angle, grain properties, and 
total opacity, $\sim$1\% is typical for models inferred for 114-426 (\idest{} 
near edge-on flared disk, \eg{} WH92 models 3,4 with $\mu < 0.3$, perhaps with 
a tenuous envelope with broad polar cavities, \eg{} WH93 models 5--8). The 
inferred intrinsic brightness of the central star is then
\filter{K}$\sim$\magnit{9}{5}, corresponding to a 1\,Myr old, 
$\sim$1.5\Msolar{} star at 450\,pc (D'Antona \& Mazzitelli 1994). Since the 
star is undetected in the 2\micron{} image to a 5$\sigma$ point source 
detection limit of \filter{K}=\magnit{18}{8}, the implied line-of-sight 
extinction towards it is $\geq$\magnit{9}{3} at 2\micron{}, or 
$A_V \geq \magap{60}$. Assuming the disk is exactly edge-on, the 
corresponding column density through the disk midplane to the central star 
is $\geq 1.1\times 10^{23}$\,cm$^{-2}$ (assuming 
1$A_V \equiv 1.9\times 10^{21}$\,cm$^{-2}$).

The disk outer radius, $r_d = 506$\,AU, is known from the data of MO96. The 
same data can be used to determine $z_d$, the scale-height at $r_d$: we fit 
a 1D average minor axis profile across the [O\,III] image of 114-426 (Fig.\,4 
of MO96) with a Gaussian characterized by Eq.\,\ref{eq:two}, yielding 
$z_d = 72$\,AU\@. Finally, we assume the disk inner radius, $r_i = 1$\,AU: 
since the surface density increases towards the center with a power $< 1$, 
our results are not too sensitive to this assumption. Integrating 
Eq.\,\ref{eq:one} (with $\alpha = -15/8$) through the midplane, we calculate 
the midplane density at the outer edge, 
$\rho_d \geq 5.8 \times 10^4$\,cm$^{-3}$. Then
integrating Eq.\,\ref{eq:three} between $r_d$ and $r_i$, the total
number of particles (HI\,+\,2H$_2$) is $\geq 3.4 \times 10^{52}$, 
yielding a mass estimate of $\geq 5.7 \times 10^{28}$\,g or $\geq 10$\Mearth.

Although it is clear that very little mass is required to render the 
central star invisible, the disk mass is likely to be significantly 
greater. Using the analytical solution for the total column density 
seen through the midplane of an edge-on disk in the $\alpha = -2 \simeq -15/8$ 
case (Section~\ref{sec:structure}), and assuming 
$\rho_d = 5.8 \times 10^4$\,cm$^{-3}$ at an outer radius of 506\,AU,
we obtain values $\sim 2\times 10^{20}$\,cm$^{-2}$ or $A_V\sim\magnit{0}{1}$
just inside the edge. Yet the disk must have a significantly higher column
density there to be {\em seen\/} as an edge: the results of MO96 show that 
the attenuation at the edge of the disk is at least 85\%, equivalent to
$A_V\geq\magap{2}$, in turn implying $\rho_d \geq 1.2\times 10^{6}$\,cm$^{-3}$.
This is a plausible value if (for example) the location of the disk edge is 
set by pressure balance with the surrounding \HII{} region: assuming
a temperature and density for the latter of 10$^4$\,K and 10$^4$\,cm$^{-3}$ 
and an outer disk temperature of 10--100\,K, densities at the disk edge of
$10^6$--$10^7$\,cm$^{-3}$ would be predicted. 

Inserting $\rho_d \geq 1.2\times 10^{6}$\,cm$^{-3}$ into Eq.\,\ref{eq:three} 
along with our other derived disk parameters yields a disk mass of 
$\geq 1.1\times 10^{30}$\,g or $\geq 5\times 10^{-4}$\Msolar. While
still considerably less than the MMSN, this is a lower limit, and the 
disk might be more massive. However, our present simple calculations based
on lower-limit extinction measurements alone cannot be pushed much further. 
Burrows \etal{} (1996) have shown for HH\,30 that more realistic mass 
estimates can be made by fitting the structure of the polar reflection 
nebulae, in particular the width of the dark lane of obscuration between 
them, and such an approach will be taken with 114-426 when our higher 
S/N and resolution Cycle~7 HST data are available. Ultimately, the most 
meaningful estimate of disk mass must come from measurements of optically 
thin tracers at millimeter wavelengths: such observations have been obtained 
for 114-426 and will be discussed in a future paper.

\section{Conclusions}
We have presented preliminary NICMOS images of the Orion~114-426 circumstellar 
disk. The disk appears to be roughly 20\% smaller at 1.87\micron{} than at
at 0.6\micron, which we attribute to a standard radial power-law density 
distribution inside a heavily truncated edge to the disk at $\sim$\,500\,AU 
radius. The general wavelength-dependent morphology of the polar lobes is 
consistent with models of a thin, massless disk surrounded by a tenuous 
envelope with broad polar cavities, and their flux was used as an indirect 
way of estimating the mass of the central star to be 1.5\Msolar. The 
non-detection of the star at 2\micron{} was then used to estimate the 
minimum total column density through the disk midplane, and thus a 
minimum disk mass of $\geq 10$\Mearth, assuming that it follows the 
standard form for a massless disk in Keplerian rotation. The hard edge 
of the disk was used to improve the lower-limit estimate under similar 
assumptions, resulting in a mass $\geq 5\times 10^{-4}$\Msolar. 

\section*{Acknowledgements}
We thank the University of Arizona, Ball Aerospace, Rockwell, NASA GSFC, 
and the STScI for developing, managing, and commissioning NICMOS\@. 
MJM thanks Abi Saha for brokering the agreement between the NICMOS IDT 
ERO and Cycle 6/7 GO teams, Chris Skinner and Eddie Bergeron for 
data reduction help, and Matthew Bate, Olaf Fischer, Bob O'Dell, Peter Schilke,
and Hans Zinnecker for useful discussions. Finally, we would like to thank the 
referee, Karl Stapelfeldt, for his insightful comments.

\newpage

\begin{figure}
\centerline{\psfig{file=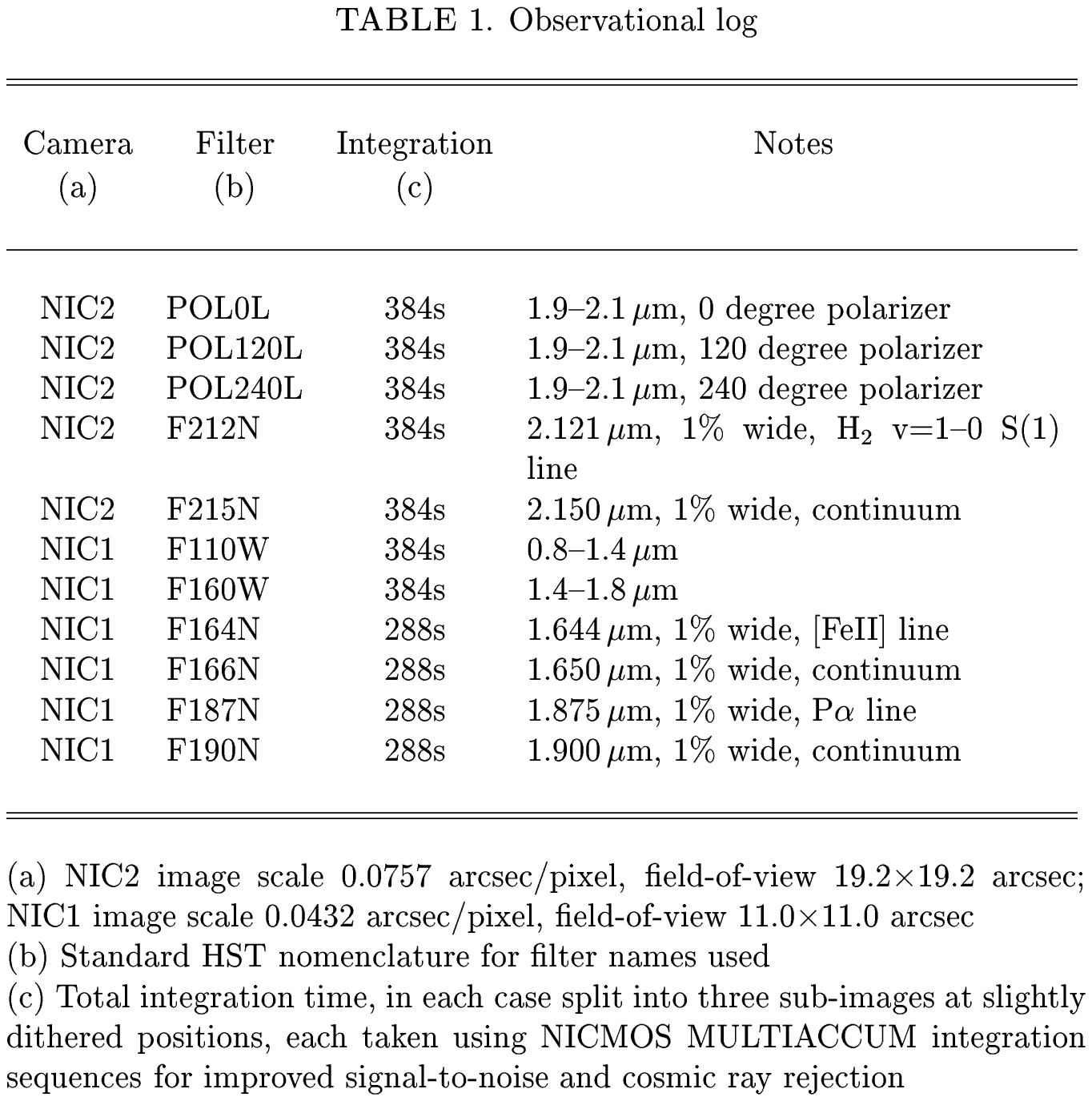}}
\end{figure}

\newpage

\begin{figure}
\centerline{\psfig{file=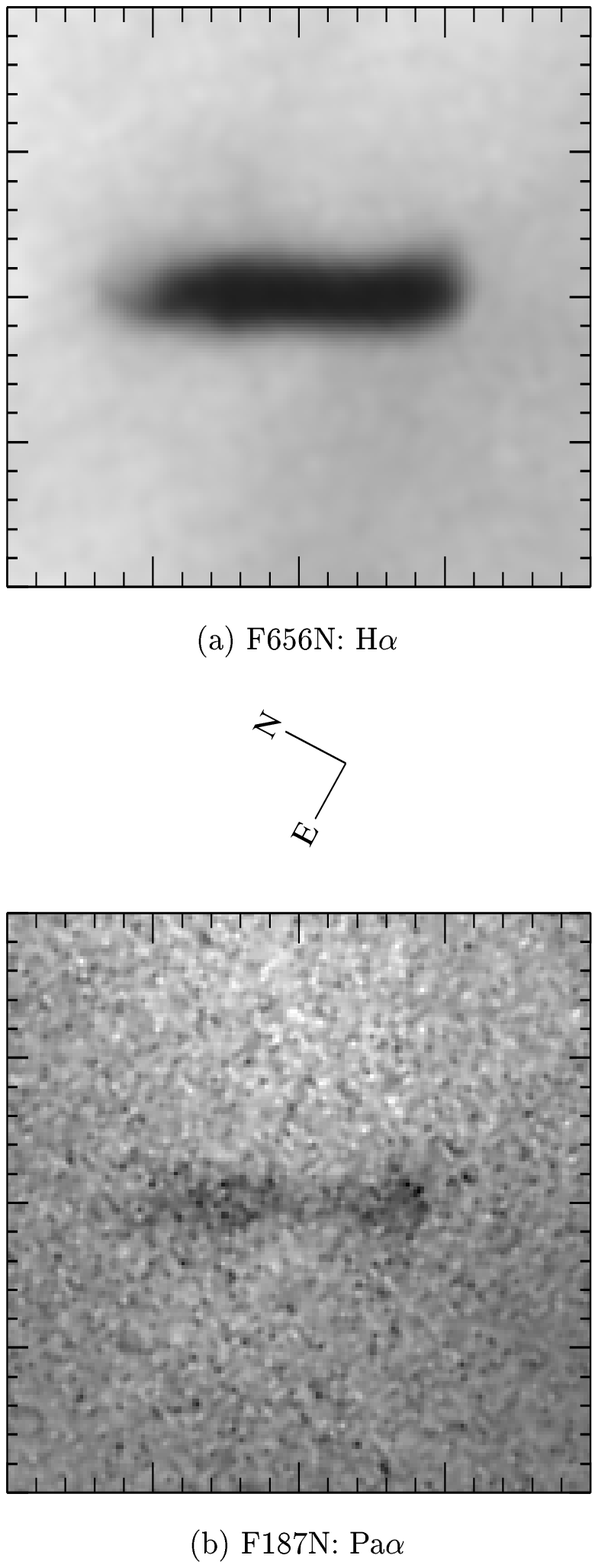,height=8in}}
\caption{%
Emission-line images of the 114-426 silhouette disk. 
(a) H$\alpha$ 6565\AA{}, (b) Pa$\alpha$ 1.87\micron. Both images have 
been magnified and rotated to a common scale and orientation. Each image 
is $4\times 4$ arcsec in size or $1800\times 1800$\,AU at 450\,pc. Major 
tick marks are at 1 arcsec intervals. The intensity scaling is arbitrary.
}
\end{figure}
\newpage

\begin{figure}
\centerline{\psfig{file=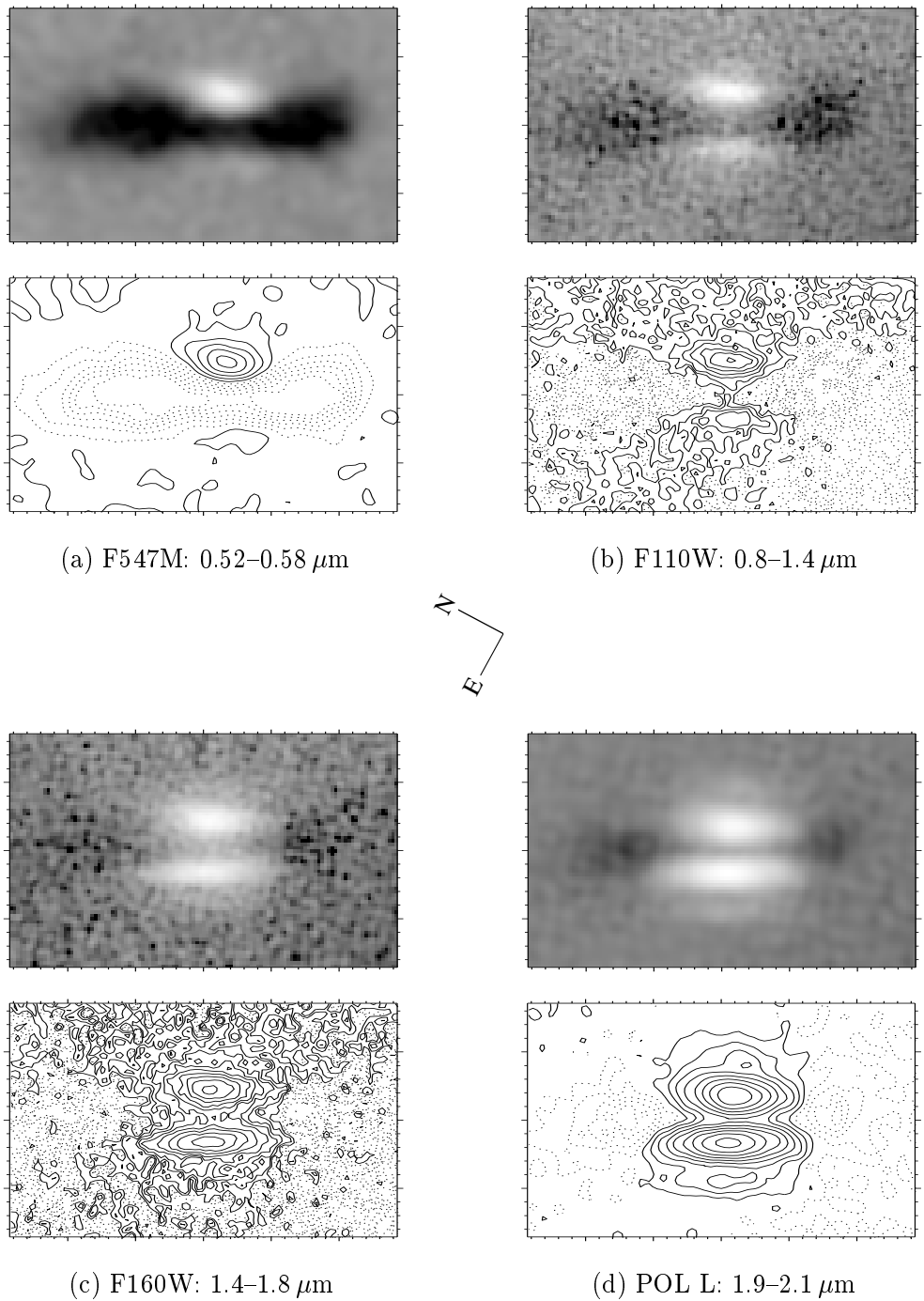,height=7.25in}}
\caption{%
Continuum images of the 114-426 disk and polar lobes
in grayscale and contour form. (a) F547M (0.52--0.58\micron), (b) F110W 
(0.8--1.4\micron), (c) F165W (1.4--1.8\micron), (d) intensity image 
derived from the POL0L/POL120L/POL240L polarization images (1.9--2.1\micron). 
All images have been magnified and rotated to a common scale and
orientation. Each image is $2.86\times 1.71$ arcsec in size or 
$\sim 1290\times 770$\,AU at 450\,pc. Major tick marks are at 1 arcsec 
intervals. The absolute intensity levels are arbitrary, but the contours
are spaced by 0.1\,dex. Contours below the local mean background 
tracing the silhouette disk are drawn as dashed lines.
Note that the fainter ``side-lobes'' seen above and below the 
main nebulae in the POL\,L image are artifacts resulting from the 
diffraction PSF of the HST+NICMOS. 
}
\end{figure}

\newpage

\begin{figure}
\centerline{\psfig{file=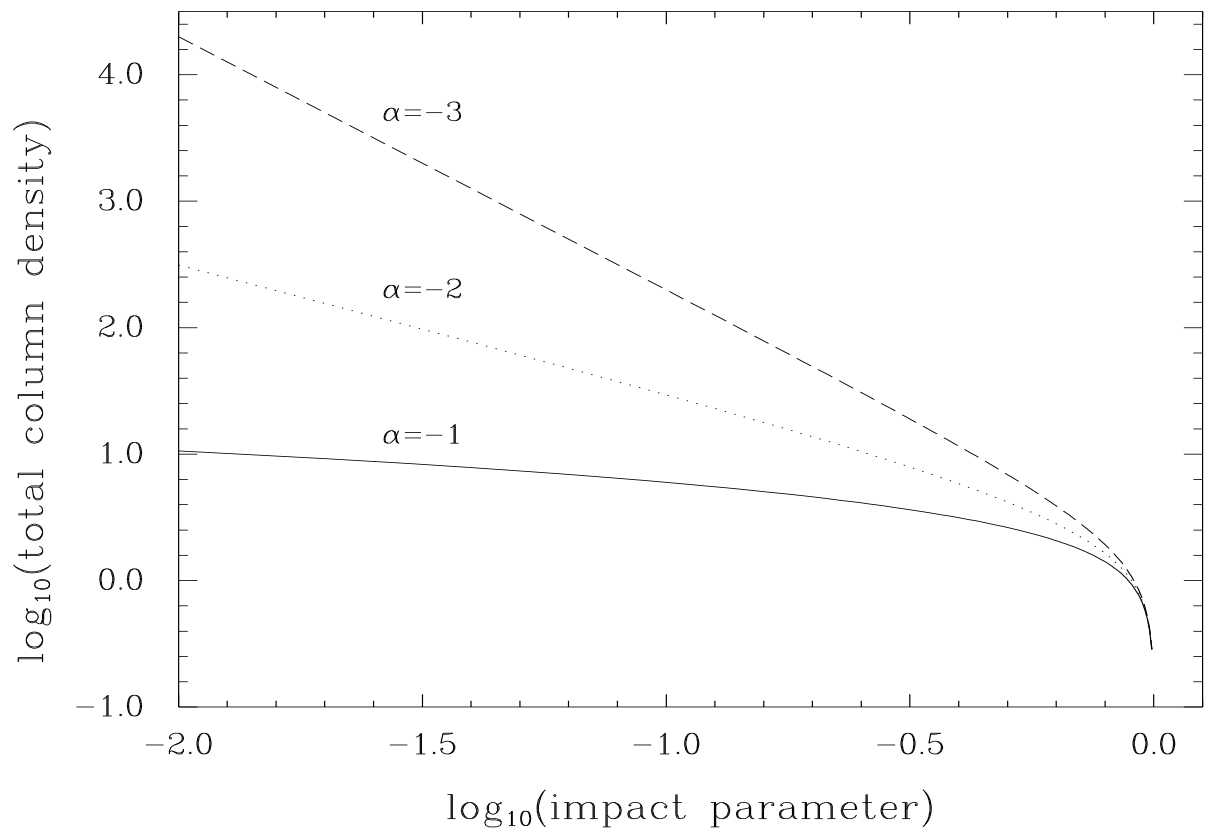,width=\textwidth}}
\caption{Model midplane column density profiles for an 
edge-on disk with an outer edge at unity radius. The total integrated
column density (in arbitrary units) is shown as a function of the
impact parameter, \idest{} the distance off-center at which the column 
is being measured. The three curves correspond to underlying radial 
power-law density distributions of the midplane density with 
$\rho \propto r^{\alpha}$, with $\alpha = -1, -2,$ and $-3$. See
text for further details.
}
\end{figure}

\vfill\eject
\end{document}